\pdfoutput=1
\documentclass[11pt]{article}
\usepackage{jheppub}

\usepackage{customprelude}

\title{\centering $\cN=3$ four dimensional field theories}

\author{I\~naki Garc\'ia-Etxebarria$^{a}$}
\author{and Diego Regalado$^{a,b}$}

\affiliation{$^{a}$Max Planck Institute for Physics, F\"ohringer Ring 6, 80805 Munich, Germany}
\affiliation{$^{b}$Institute for Theoretical Physics and
Center for Extreme Matter and Emergent Phenomena,\\
Utrecht University, Leuvenlaan 4, 3584 CE Utrecht, The Netherlands}

\emailAdd{inaki@mpp.mpg.de}
\emailAdd{regalado@mpp.mpg.de}

\abstract{We introduce a class of four dimensional field theories
  constructed by quotienting ordinary $\cN=4$ $U(N)$ SYM by particular
  combinations of R-symmetry and $SL(2,\bZ)$ automorphisms. These
  theories appear naturally on the worldvolume of D3 branes probing
  terminal singularities in F-theory, where they can be thought of as
  non-perturbative generalizations of the O3 plane. We focus on cases
  preserving only 12 supercharges, where the quotient gives rise to
  theories with coupling fixed at a value of order one. These
  constructions possess an unconventional large $N$ limit described by
  a non-trivial F-theory fibration with base
  $AdS_5\times (S^5/\bZ_k)$. Upon reduction on a circle the $\cN=3$
  theories flow to well-known $\cN=6$ ABJM theories.}

\begin{document}


\makeatletter
\let\old@fpheader\@fpheader
\renewcommand{\@fpheader}{\old@fpheader\hfill
MPP-2015-307}
\makeatother

\maketitle
\newpage

\section{Introduction}

One of the fascinating properties of string theory lies in its ability
to geometrize various deep and subtle field theory phenomena, often
giving insight not available in any other known way.

A particularly fruitful geometric context in which to study four
dimensional field theories is F-theory \cite{Vafa:1996xn}. It can be
understood as a geometric rewriting of IIB string theory with an
axio-dilaton varying over a base, but crucially also as a particular
limit of M-theory on the Calabi-Yau fourfold defined by the
axio-dilaton fibration. Questions about the field theory translate
into questions about the geometry of the fourfold, and can often be
dealt with using algebraic geometry methods.

A crucial point is that most of the interesting phenomena (for a field
theorist using M-theory as a computing tool) arise when the geometry
develops singularities. Much of the recent work dealing with F-theory
model building (starting with
\cite{Donagi:2008ca,Beasley:2008dc,Hayashi:2008ba,Beasley:2008kw}),
for instance, is concerned with the construction of appropriately
singular geometries in order to model features of the standard model,
and of its supersymmetric and grand unified extensions.

While the physics of interest happens on the singular locus, we do not
have much control directly over M-theory on singular spaces, so in
practice one constructs a family of smooth Calabi-Yau spaces
parameterized by some parameter $t$, such that when $t\to 0$ the
Calabi-Yau develops the singularity of interest, but the
representatives for $t\neq 0$ are all smooth. By following which
cycles in the geometry vanish as we approach $t=0$ we can read off
which BPS states become massless at the singular point
\cite{Witten:1996qb}, and thus reconstruct much of the low energy
physics.

\medskip

In this note we deal with singularities in the M-theory description of
certain four dimensional theories that cannot be approached in this
way: one cannot construct a family of smooth Calabi-Yau spaces
abutting the singular Calabi-Yau. Nevertheless, as we will discuss
below, examples of singularities in the M-theory fourfold that cannot
be smoothed out in a supersymmetric way are both rather common --- the
ordinary O3 plane gives rise to such a structure, and the local
geometry of the M-theory backgrounds that appear in our construction
is precisely the one in the ABJM construction
\cite{Aharony:2008ug,Aharony:2008gk} --- and quite interesting from a
field theoretic point of view. We focus on the simplest hitherto
unknown examples, which happen to preserve twelve (but not sixteen)
supercharges in four dimensions, which we refer to as $\cN=3$ theories
in the customary fashion. It is well known that four dimensional
supergravity Lagrangians preserving just twelve supercharges exist,
but to our knowledge this is the first time that $\cN=3$ examples
outside that class have been constructed.

The theories under analysis here have a number of amusing
properties. The most characteristic one --- and the one that allows
them to evade well known results \cite{Weinberg:2000cr} stating that
perturbative $\cN=3$ theories necessarily have $\cN=4$ supersymmetry
--- is that their associated SCFTs do not have a marginal deformation
associated with taking the gauge coupling to a perturbative regime.

This fact will in fact appear rather naturally from the construction:
when formulated in field theory terms the $\cN=3$ theories in this
paper arise as quotients of ordinary $\cN=4$ $U(N)$ SYM theory by a
symmetry involving a non-trivial action of the $SL(2,\bZ)$ duality
group, together with an action of the $SU(4)$ R-symmetry
group.\footnote{A related construction, already well studied in the
  literature \cite{Ganor:2008hd,Ganor:2010md,Ganor:2012mu} (see also
  \cite{Dabholkar:2002sy,Hull:2003kr,CatalOzer:2006mn}), considers the
  $S^1$ compactification of the four dimensional $\cN=4$ theory
  twisted by the combined action of R-symmetry, S-duality, and a shift
  along the $S^1$.} (Quotients by a subgroup of the $SU(4)$ R-symmetry
group can be seen to describe the theory of D3 branes at Calabi-Yau
orbifold singularities in perturbative IIB string theory
\cite{Lawrence:1998ja}, our work can be thought of as a generalization
of this viewpoint to F-theory, where extending the orbifold action to
the fiber is natural to consider.)  For generic values of the
Yang-Mills coupling the $SL(2,\bZ)$ duality group is not a symmetry of
the theory, and it only becomes so for specific self-dual values of
the coupling. Deformations away from this point should then be
projected out from the quotient, and indeed we show in \S\ref{sec:CPT}
that this is the case.

This construction is most naturally motivated from the M-theory
viewpoint, where our construction comes from orbifold actions on
$\bR^3\times \bC^3 \times T^2$ that only make sense for specific
values of the complex structure of the torus fiber. This was in fact
our original motivation for approaching this problem: as reviewed in
\S\ref{sec:O3} the ordinary O3 plane in IIB string theory can be
understood as an orbifold of IIB theory in flat space by an orbifold
generator involving the action of the $-1\in SL(2,\bZ)$ element of the
duality group. It is fairly natural to ask if this can be generalized
in the F-theory context to orbifolds involving duality elements acting
non-trivially on the IIB axio-dilaton, and if so what is the physics
of probe D3 branes on top of the resulting singularity. The purpose of
this note is to answer the existence question in the affirmative, and
initiate the study of the resulting field theories.

\medskip

\emph{Note added:} As we were finishing the contents of this note
\cite{Aharony:2015oyb} appeared, giving a field theoretical derivation
of some properties of $\cN=3$ SCFTs, under the assumption that such
theories exist. The properties they find seem to be in agreement with
those of the theories we construct below.

\section{D3 branes on the O3 plane}

\label{sec:O3}

In order to motivate our construction, we now revisit a familiar
system, the theory of D3 branes on top of an O3 plane. This system is
most commonly studied from the point of view of the worldsheet CFT, a
construction which we briefly review now (for a more exhaustive review
see for instance \cite{Polchinski:1998rr}). In this context the theory
on the D3 branes on top of O3 planes is defined as the quotient of the
theory of open strings moving in flat space with D3 boundary
conditions, by the orientifold action $\cI(-1)^{F_L}\Omega$. Here
$\Omega$ reverses the orientation of the worldsheet, $(-1)^{F_L}$ acts
as $-1$ on RNS and RR states, and $\cI$ acts as reflection on the
three complex directions transverse to the O3
\begin{equation}
  \cI\colon (z_1,z_2,z_3) \to (-z_1,-z_2,-z_3)
\end{equation}
leaving the four real dimensions along the O3 plane invariant. Before
quotienting by the orientifold action the low energy description of
the system is given by four dimensional $\cN=4$ $U(N)$ theory
(forgetting about the ten dimensional dynamics in the bulk, which
decouples at low energies), arising from open strings with ends on the
D3 stack. The orientifold preserves all the supersymmetry of the
original D3 stack, but projects down the gauge group to a
subgroup. For concreteness we locate the D3 stack on top of the O3
plane, i.e. at $z_1=z_2=z_3=0$. The nature of the preserved subgroup
depends on the choice of the representation of the orientifold action
on the Chan-Paton factors. There are two inequivalent choices for this
representation: we either end up with the algebra $\fso(N)$ (we will
discuss the global form of the gauge group momentarily), or
$\fsp(N)$. In this latter case we need to restrict to $N\in 2\bZ$ for
consistency.

\medskip

We now want to discuss this construction from two alternative
viewpoints, in order to motivate the generalization presented in the
next section. Consider first the description of the system in
F-theory, obtained by taking the zero size limit of the fiber for an
M-theory compactification down to three dimensions on a torus fibered
Calabi-Yau fourfold. The basic necessary fact for describing the O3 in
this language is that $(-1)^{F_L}\Omega$ lifts to an inversion of the
torus fiber, i.e. a monodromy matrix
\begin{equation}
  \cM_{(-1)^{F_L}\Omega} = \begin{pmatrix}
    -1 & 0 \\
    0 & -1
  \end{pmatrix}\, .
\end{equation}
A simple derivation of this fact can be obtained by looking to the
action of $(-1)^{F_L}\Omega$ on the IIB spacetime fields. For
instance, in the CFT language it is easy to see that both the NSNS
two-form $B_2$ and the RR two-form $C_2$ get an intrinsic minus sign
under $(-1)^{F_L}\Omega$, while in the F/M-theory language they come
from the reduction of $C_3$ along two independent one-cycles of the torus
fiber. We immediately conclude that $(-1)^{F_L}\Omega$ acts as
inversion of the torus.

The F-theory lift of a stack of $N$ D3 branes in flat space (the
$U(N)$ theory) is given by a stack of $N$ M2 branes on
$\bC^3\times T^2$. The fibration is trivial in this case, with the
complex structure of the torus arbitrary. This arbitrariness maps to
the existence of the marginal deformation of $\cN=4$ $U(N)$ changing
the value of the complexified coupling. The orientifolded system can
then be constructed in F-theory by taking the quotient
\begin{equation}
  \sigma\colon (z_1,z_2,z_3,u) \to (-z_1, -z_2, -z_3, -u)
\end{equation}
with $u$ the flat coordinate of the $T^2$. That is, $T^2$ is the
quotient of $\bC$, parameterized by $u$, by some lattice
$\bL=\{ae_1 + b e_2\}$ with $a,b\in \bZ$. This involution clearly
exists for any complex structure of the torus, since a change of sign
maps any integer lattice $\bL$ to itself.

The resulting geometry $X=(\bC^3\times T^2)/\sigma$ has four fixed
points at $z_1=z_2=z_3=0$ and $u=-u$ mod $\bL$. (For instance, if we
take $e_1=1,e_2=\tau$, the fixed points are at
$\{0,\frac{1}{2}, \frac{1}{2}\tau, \frac{1}{2}(\tau+1)\}$.) Close to
any of these fixed points we have a $\bC^4/\bZ_2$ geometry, with the
$\bZ_2$ inverting all coordinates of the $\bC^4$. Notice that the
$\bC^4/\bZ_2$ singularity is terminal \cite{morrisonStevens,Anno}: it
admits no supersymmetric resolution or deformation into a smooth
fourfold. This is in good agreement with the fact that there are no
twisted sectors that could smooth out the O3 plane.

\medskip

We will come back to the F/M description momentarily, but let us first
consider the field theory description of the orientifold
operation. Such a description in terms of the low energy EFT must
exist, since the $\cN=4$ theories before and after orientifolding are
consistent truncations of the full string theory. It will be
illuminating to consider the simplest case: the theory on one mobile
D3 brane on top of an O3$^-$ plane. As is well known, this is given by
the $\cN=4$ theory with $\fso(2)$ algebra. At this level the theory is
identical to the $U(1)$ theory arising from a D3 on flat space,
without considering the orientifold. There is a difference when one
considers the global structure of the gauge group, though. This is
perhaps best seen by comparing the moduli spaces. For the $U(1)$
theory we expect the moduli space to be simply $\bC^3$, while for the
$\fso(2)$ theory we would rather expect $\bC^3/\bZ_2$ (with the
$\bZ_2$ acting with a minus sign on all $\bC^3$ coordinates). The
moduli space arises from the vevs of the scalars on the $\cN=4$ vector
multiplet, which take values in the adjoint. A way to achieve our
goal, while keeping supersymmetry, is then to take a quotient by a
$\bZ_2$ acting as $t\to -t$ on the algebra generators. For the cases
of orientifolds giving rise to $SO(2N+1)$ and $\Sp(2N)$ groups the
expected $\bZ_2$ quotient of moduli space is always present as part of
the Weyl group, but the Weyl group is trivial for $\fso(2)$. Rather,
the sign action is precisely the $\bZ_2$ outer automorphism of the
$SO(2)$ group enhancing it to $O(2)$, so we learn that the expected
form of the gauge group is $\bZ_2\ltimes SO(2)=O(2)$.\footnote{This
  conclusion for the global form of the gauge group can be reached in
  many additional ways. For example, upon compactification of the
  theory on a circle, and T-dualization, the global form must be
  $O(2)$ so that we can construct the component containing two
  $\widetilde{\text{O2}}^-$ planes. Or simply by looking to the
  perturbative symmetry group: as opposed to the type I case, in this
  case there are no finite action non-BPS instantons spoiling the
  conclusion.\label{fn:O(2)}}

\medskip

There is a way of understanding this $\bZ_2$ automorphism that
connects nicely to the F/M description, and which immediately suggests
the generalization that is the main topic of this note. The $\cN=4$
theory with gauge group $U(1)$ has a global symmetry which includes the
R-symmetry group $SU(4)_R$, that has a natural interpretation as
the rotation group in the $\bR^6$ orthogonal to the stack. We wish to
focus on the element $r\in SU(4)_R$ acting as
\begin{equation}
  r\colon (A_\mu,\lambda_\alpha^a,\phi^I)\to (A_\mu,\sqrt{-1}\,\lambda_\alpha^a,
  -\phi^I)
\end{equation}
on the components of the $\cN=4$ vector multiplet, corresponding to
$SO(6)$ rotations in the $\bR^6$ acting as an overall sign
$z_i\to -z_i$. Here $a$ and $I$ are indices in the spinor and vector
representations of $SU(4)_R$. This action does not preserve
supersymmetry by itself, in agreement with the fact that $\bC^3/\bZ_2$
is not a Calabi-Yau space. Notice also that $r^2\neq 1$.

There is, in addition, a subgroup of the $SL(2,\bZ)$ duality group
that acts nontrivially as a symmetry of the theory for any value of
the coupling $\tau$, it is given by $s=-1\in SL(2,\bZ)$. It leaves
$\tau$ invariant, but it acts non-trivially on the space of states. In
particular, it sends a BPS state with electric and magnetic charges
$(p,q)$ to another state with charges $(-p,-q)$, so it acts with a
minus sign on the gauge boson $A_\mu$. The full action of $s$ on the
components of the $\cN=4$ multiplet is
\begin{equation}
  s\colon (A_\mu,\lambda_\alpha^a,\phi^I)\to (-A_\mu,\sqrt{-1}\,\lambda_\alpha^a,
  \phi^I)\, .
\end{equation}
(For a field theory derivation of the action on the gauginos see for
example \cite{Kapustin:2006pk}.) Notice again that $s^2\neq 1$. The
inversion of the $(p,q)$ charges has a natural interpretation as the
$\bZ_2$ action on the F-theory torus, since light strings come from M2
branes wrapped on one-cycles.

We now have an intrinsic definition of the $\bZ_2$ orientifold action
in the field theory: it is simply given by the quotient by $r\cdot s$,
which acts as a global $-1$ on the whole vector multiplet:
\begin{equation}
  \label{eq:O(2)-orientifold-quotient}
  O(2)_{\cN=4} = \frac{U(1)_{\cN=4}}{r\cdot s}\, .
\end{equation}

One may worry about the self-duality of the O3$^-$ under S-duality,
given that the gauge group is $O(2N)$ rather than $SO(2N)$. We
postpone a discussion of this point to appendix \ref{ap:self}.

\medskip

\subsubsection*{Interacting theories}

Consider the $O(2N)$ theory, with $N>1$. In order to construct this
theory field theoretically, starting with a theory without
orientifold, we need to
generalize~\eqref{eq:O(2)-orientifold-quotient} slightly. The first
way of doing this is familiar from the worldsheet description of the
orientifold. Start with the theory in the double cover of the geometry
and in the absence of orientifold, having twice the amount of mobile
branes, giving rise to a gauge group $U(2N)$. In this doubled theory,
in addition to specifying the $r\cdot s$ quotient, we specify a
Chan-Paton action on the gauge bosons
\begin{equation}
  A \to -A^t
\end{equation}
in such a way that only antisymmetric gauge matrices survive, giving
rise to the $\fso(2N)$ gauge algebra.

Nevertheless, for the non-perturbative generalizations that we have in
mind a different description is perhaps more adequate. We work
directly on the quotient geometry, where we have $N$ mobile
branes. Away from the singularity in the geometry, the low energy
gauge group is $U(N)$. As the stack of branes hits the singularity
some light modes become massless, and enhance the algebra to
$\fso(2N)$. In the M-theory description these light modes arise from
BIon excitations on the probe M2 brane stack. More precisely, the
massless modes transform in the $\asymm \oplus \ov{\asymm}$
representation of the $U(N)$ group. Let us separate slightly the M2
branes in the stack for ease of visualization, and consider the
topology of the quotient geometry, removing the origin of the base
$\bC^3$ for simplicity. The projection of the orbifold to the base has
topology $\bR\times (S^5/\bZ_2)$, which for the purposes of homology
is the same as $S^5/\bZ_2=\bR\bP^5$. There is a non-trivial cycle in
this $\bR\bP^5$ arising in its $S^5$ double cover from the path
between one point and its image. The action on the $T^2$ fiber
homology as we go around this closed path is changing the sign of the
generators: a $(p,q)$ cycle goes to $(-p,-q)$. So, between any two
distinct M2 branes on the stack, generating a Cartan subgroup
$U(1)^2$, there are two non-trivial paths which differ by the
non-trivial torsional element in the base. Due to the non-trivial
monodromy in the fiber, if one gives a state with $(1,1)$ charges
under the $U(1)^2$ Cartan, the other gives a state with $(1,-1)$
charges. We recognize these states as the generators of the adjoint
and antisymmetric representations of $U(N)$, and together they enhance
$U(1)^N$ to $O(2N)$.

We could have also considered states going from one brane on the stack
to itself, wrapping the non-trivial cycle in $\bR\bP^5$. Whether they
exist or not depends on the NSNS charge of the O3 plane: they are
absent for O3$^-$ and they are present for O3$^+$. Had we included
them, the antisymmetric would have been replaced by a symmetric
representation, giving rise to an enhancement to $\Sp(2N)$ when the
stack hits the singularity. How do we explain this distinction in the
resulting projection without resorting to CFT language? One suggestive
observation is that the O3$^-$ and O3$^+$ can be transformed into each
other by dropping a NS5 wrapped on the nontrivial (twisted) two-cycle
of $\bR\bP^5$ \cite{Witten:1998xy,Hyakutake:2000mr}. The NS5 brane and
the fundamental string are electric-magnetic duals, so this suggests
that the origin of the distinction, from the target space point of
view, may come from an unsatisfied Dirac quantization condition on the
F1/M2 worldvolume if we drop an even number of NS5/M5 branes on the
O3$^-$ plane. It would be very interesting to develop this viewpoint
further as a possible tool in order to study the spectrum of BPS
states in the more exotic configurations we construct later in the
paper.

\subsubsection*{Flux classification of O3 types}

\begin{figure}
  \centering
  \includegraphics[width=0.4\textwidth]{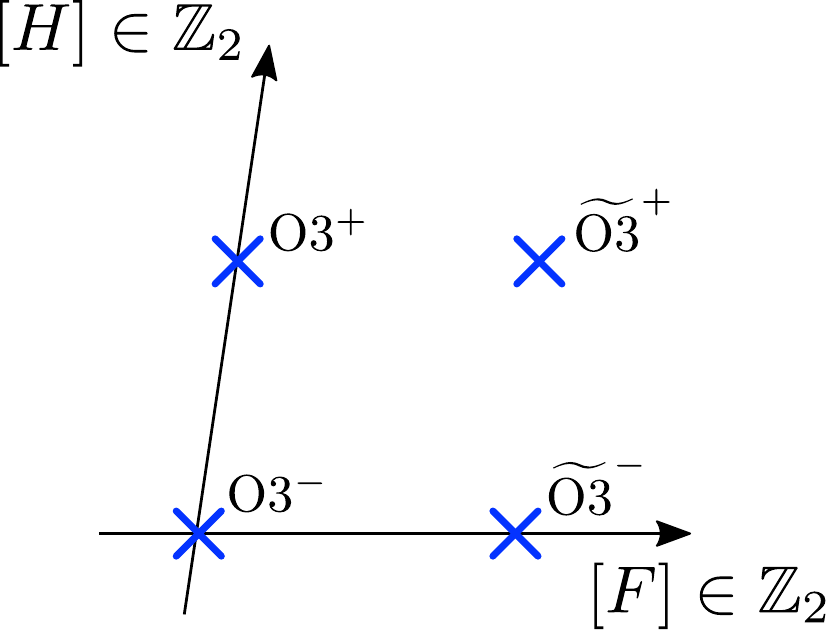}
  \caption{The four types of O3 planes, classified based on their
    discrete torsion \cite{Witten:1998xy}. A positive superscript
    denotes non-trivial NSNS torsion, while a tilde denotes nontrivial
    RR torsion.}
  \label{fig:O3planes}
\end{figure}

It is a familiar fact that there are various kinds of O3 planes, as was
implicit in the discussion above. A convenient classification is in
terms of the discrete fluxes $[H]$ and $[F]$ for the NSNS and RR
two-forms around the O3 plane \cite{Witten:1998xy}. Both kinds of
torsion are $H^3(S^5/\bZ_2,\widetilde{\bZ})=\bZ_2$ valued, so we have
four distinct possibilities, shown in figure~\ref{fig:O3planes}
together with conventional names for each orientifold type. We
emphasize that the statement that we have four different kinds of O3
planes in IIB is perturbative: once we take the $SL(2,\bZ)$ duality of
the theory into account the four orientifolds organize themselves into
a singlet of $SL(2,\bZ)$ (the O3$^-$) and a triplet (the
$\widetilde{\text{O3}}^-$, O3$^+$ and
$\widetilde{\text{O3}}^+$). Accordingly, we expect that there is no
distinction between the M-theory lifts of the orientifolds in the
triplet, but that they are distinct from the M-theory lift of the
O3$^-$.

The discrete fluxes $[H]$ and $[F]$ lift to the M-theory description
as discrete fluxes for the $C_3$ form. More precisely, around each
$\bC^4/\bZ_2$ local singularity in the M-theory dual of the O3 plane,
we can turn on a nontrivial $C_3$ represented by the torsional
generator of $H^4(S^7/\bZ_2,\bZ)=\bZ_2$.

Alternatively \cite{Witten:1998xy,Hyakutake:2000mr}, we can start with
the O3$^-$ plane, and generate the discrete $[H]$ and $[F]$ torsions
by dropping an NS5 or D5 wrapped on the non-trivial $\bZ_2$ generator of
$H_2(S^5/\bZ_2,\widetilde{\bZ})$. In the M-theory language, we change
the flux on the $\bC^4/\bZ_2$ singularity by dropping a M5 wrapped on
the non-trivial generator of $H_3(S^7/\bZ_2)=\bZ_2$.

In either description, the net result is that we have $2^4=16$
discrete choices for the discrete torsion on the M-theory lift of the
orientifold. (There are actually only five inequivalent configurations
in M-theory, once we account for possible torus redefinitions.) The
fact that we have a larger number of discrete choices in M-theory than
in IIB reflects the fact that some of these configurations can become
trivial or equivalent once we take the F-theory limit
\cite{Hanany:2000fq}.

This is most easily seen in the current case by performing an intermediate
step in the duality between IIB and M-theory: in IIA string theory ---
obtained by reducing M-theory on one of the cycles of the $T^2$, or
T-dualizing on the circle in which we reduce IIB --- we also have four
types of O2 planes, distinguished by the flux of the NSNS three-form
and the flux of the RR four-form. We denote the different planes by
O2$^\pm$ and $\widetilde{\text{O2}}^\pm$ in analogy with the O3
case. The lift of the different O2 planes to M-theory is given by
M-theory on $(\bR^7\times S^1)/\bZ_2$, with the $\bZ_2$ acting as a
reflection on all directions. The reflections on the $\bR^7$ base are
inherited from IIA, while the fact that the M-theory $S^1$ gets
reflected can be understood from the fact that D0 branes transform
with an intrinsic minus sign under a O2, and D0 branes lift to
momentum along the M-theory circle. Over the origin in $\bR^7$ where
the O2 is located the M-theory $S^1$ degenerates to a segment, and the
local geometry around the endpoints of the segments is
$\bC^4/\bZ_2$. For future convenience we denote the $\bC^4/\bZ_2$
geometry with $f=\{0,1\}$ units of flux by $\OM_2^f$. Thus, there
should be a one-to-one map between types of O2 planes and pairs of
$\OM$ planes. The map can be easily worked out by computing the
charges under $C_3$, see for instance \cite{Hanany:2000fq}, with the
result displayed in table~\ref{table:O2-lifts}.

\begin{table}
  \centering
  \begin{tabular}{c|c|c|c}
    O2$^-$ & $\widetilde{\text{O2}}^-$ & O2$^+$ & $\widetilde{\text{O2}}^+$
    \\
    \hline
    (0,0) & (1,1) & (1,0) & (0,1)
  \end{tabular}
  \caption{Lift to M-theory of the different types of O2 planes in
    IIA. The labels in the bottom row denote the discrete flux on the
    two $\OM_2$ planes present in the lift, with 0 denoting trivial
    flux.}
  \label{table:O2-lifts}
\end{table}

We now come back to the issue of equivalence of M-theory
configurations under the F-theory limit. Consider for example the case
of one mobile D3 brane probing a O3$^-$ plane. The low energy dynamics
is described by a $O(2)$ group. There are two components in moduli
space once we reduce on a circle, depending on the Wilson line along
the circle: in one we have (after T-duality along the circle) two
O2$^-$ planes and a mobile D2, and in the other we have two
$\widetilde{{\rm O}2}^-$ planes with no mobile brane. The M-theory
lift of these two configurations is different, but the distinction
disappears in the F-theory limit, where the Wilson line becomes
irrelevant. One can also see in this way that two of the other
orientifolds give rise to shift orientifolds in IIB acting on the
compactification circle, and thus become trivial orientifolds in the
strict F-theory limit.

\medskip

These remarks complete our discussion of the O3 plane. As we see,
while the object was originally introduced in CFT language, it can be
described intrinsically both in M-theory and in effective field theory
(although it would be good to develop both of these viewpoints
further). Unlike the CFT viewpoint, these
generalize to non-classical configurations, so we now proceed to
discuss some of these generalizations.

\section{F-theory at $\cN_{4d}=3$  singularities}

\label{sec:F-theory}

We will be interested in the F-theory limit of M-theory on abelian
orbifolds of the form $\bR^{1,2}\times (\bC^3\times T^2)/\bZ_k$ where
$k\in\{2,3,4,6\}$. The action of $\bZ_k$ on the complex coordinates $(z^1,z^2,z^3,z^4)$ of $\bC^3\times T^2$ is given by\footnote{This convention is such that $\bZ_k$ lies in $SU(4)$, which makes the orbifold a Calabi-Yau. The choice $z^i\rightarrow e^{2\pi i/k}z^i$ is physically equivalent \cite{Aharony:2008ug} but we find our convention more convenient for our purposes.}
\begin{equation}\label{orbi}
z^i\longrightarrow e^{2\pi i v_k^i}z^i
\end{equation}
with $v_k=(1,-1,1,-1)/k$ and where $z^4=x+\tau y$ with $\tau$ the
complex structure of $T^2$. These values for $k$ are the only
possibilities that are well-defined on the torus.\footnote{See
  \cite{morrisonStevens,Anno} for classifications of codimension four
  terminal Gorenstein quotient singularities. The cases we consider
  fall into this classification so they are terminal.} The case $k=2$
corresponds to the O3 plane studied in the last section (and which
preserves sixteen supercharges) so we will focus on the other
possibilities, which preserve just twelve out of the sixteen
supercharges. (A proof of this fact will be given in
\S\ref{sec:supercharges}.)

\begin{table} 
\renewcommand{\arraystretch}{1.25}
\setlength{\tabcolsep}{5pt}
\begin{center}
  \begin{tabular}{c||c|c|c|c}
    & $\tau$ & $\chi$ & $Q(\OM_{k,0})$ & $Q(\OF_{k,0})$ \\\hline\hline
    $k=2$ & any & 384 & $-1/16$ & $-1/4$ \\\hline
    $k=3$ & $\exp(\pi i/3)$ & 216 & $-1/9$ & $-1/3$ \\\hline
    $k=4$& $i$ & 240 & $-5/32$ & $-3/8$ \\\hline
    $k=6$& $\exp(\pi i/3)$ & 240 & $-35/144$ & $-5/12$
  \end{tabular}
\end{center}
\caption{Data for the $\cN\geq 3$ orientifolds analyzed in this
  paper. D3 charges for the $\OF_{k,0}$ planes are given in units of
  mobile D3 branes.}
\label{kas}
\end{table}

Notice that the $\bZ_k$ symmetry can only act as an involution of the
torus for certain fixed values of the complex structure $\tau$, which
we list in table~\ref{kas}. Since the complex structure of the torus
in M-theory corresponds to the axio-dilaton when we take the F-theory
limit, these backgrounds are intrinsically non-perturbative in ten
dimensions. The situation is similar to F-theory compactifications on
certain singular limits of K3 \cite{Dasgupta:1996ij} or elliptically
fibered Calabi-Yau threefolds with Hirzebruch bases
\cite{Witten:1996qb}. However, in those cases one can smoothly move in
complex structure moduli space of the K3 to a perturbative
configuration. As emphasized in the introduction, this is not possible
for the orbifolds we are considering.

These orbifolds, or generalized orientifolds, can also be understood
in field theory. Recall that the R-symmetry in the worldvolume of a
stack of D3 branes can be understood as rotations on the six
directions transverse to the worldvolume, while $SL(2,\bZ)$ duality is
associated with large diffeomorphisms of the torus fiber in the
F-theory description. When we take the $(\bC^3\times T^2)/\bZ_k$
quotient in~\eqref{orbi}, we are taking a quotient of $\cN=4$ by the
combined action of a R-symmetry generator $r_k$, and an $SL(2,\bZ)$
symmetry generator $s_k$. With the inclusion of appropriate massless
sectors (which we do not describe in this paper, but follow from the
M-theory description, in analogy with the extra massless two-index
tensors in the O3 case described above), this defines a theory by the
analog of formula~\eqref{eq:O(2)-orientifold-quotient}. Our goal in
the rest of the note will be to describe some properties of this
theory.

\subsection{Orientifold variants}

\begin{figure}
  \centering
  \includegraphics[width=0.6\textwidth]{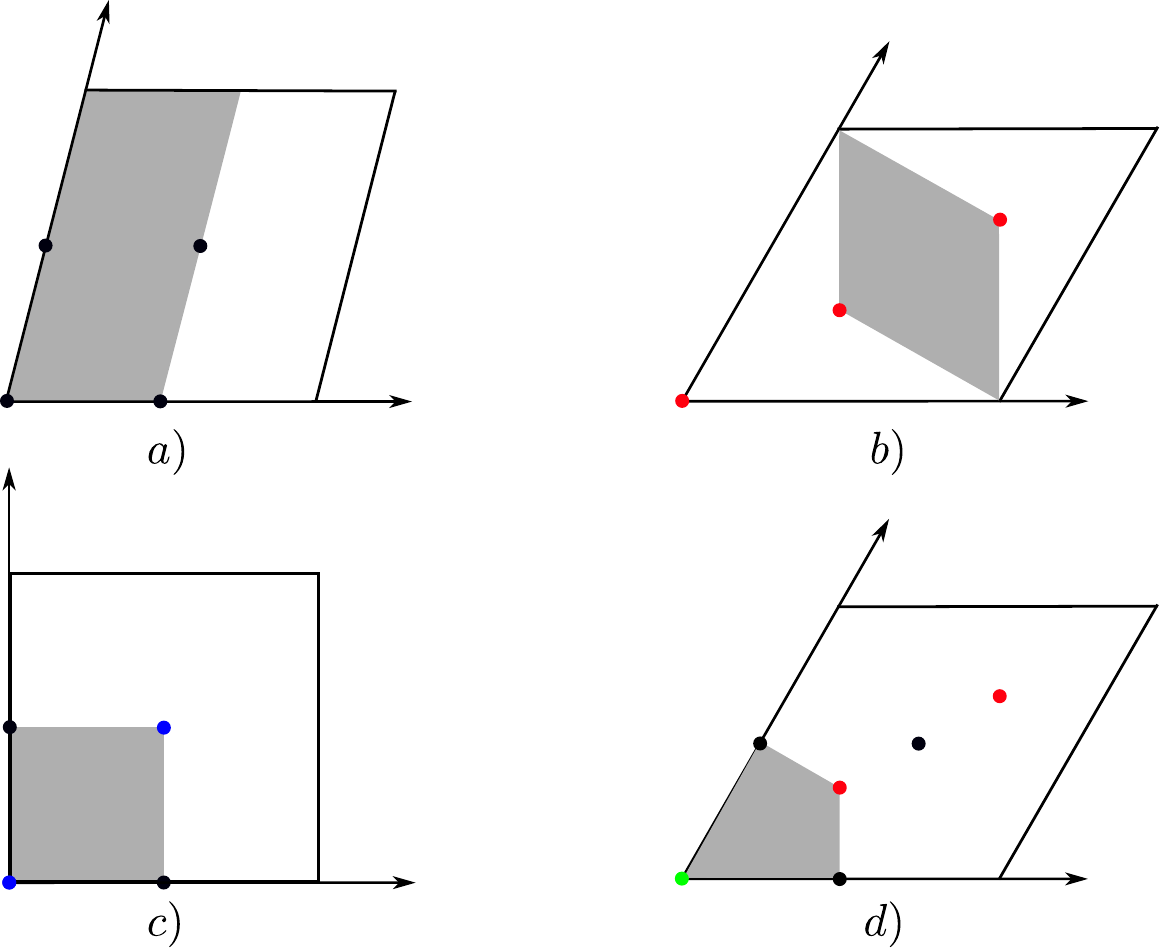}
  \caption{$T^2/\bZ_k$ for $a)\,k=2$, $b)\,k=3$, $c)\,k=4$ and $d)\,k=6$. The black, red, blue and green dots represent the $\bZ_k$ fixed points for $k=2,3,4$ and $6$ respectively. The grey regions denote the fundamental domains in each case.}
  \label{fig:tori}
\end{figure}

A basic quantity to compute for any BPS object in string theory is its
charge. In the M-theory formulation the contribution to the charge
under $C_3$ of a $\bC^4/\bZ_k$ singularity comes from the curvature
coupling~$-\int C_3\wedge I_8(R)$, where $\int I_8(R)=\chi/24$. For
the M-theory configurations of interest to us in this paper, we
typically have more than one fixed point. More precisely, the topology
of the singular fiber at the origin of the $\bC^3$ base is of the form
displayed in figure~\ref{fig:tori} (see for example
\cite{Dulat:2000xj,Nilse:2006jv} for details), namely
\begin{itemize}
\item $T^2/\mathbb Z_2$ is a sphere with four $\mathbb Z_2$ orbifold
  points.
\item $T^2/\mathbb Z_3$ is a sphere with three $\mathbb Z_3$ orbifold
  points.
\item $T^2/\mathbb Z_4$ is a sphere with two $\mathbb Z_4$ and one
  $\mathbb Z_2$ orbifold points.
\item $T^2/\mathbb Z_6$ is a sphere with a $\mathbb Z_6$,
  $\mathbb Z_3$ and $\mathbb Z_2$ orbifold points.
\end{itemize}
In the F-theory limit, the D3 charge of the 4d object $\OF_{k,0}$ (the
limit of M-theory on $\bR^{1,2}\times(\bC^3\times T^2)/\bZ_k$, with no
torsional flux) is given by the sum of the contribution of each fixed
point.

The charge under the M-theory three-form $C_3$ of the orbifold
$\bC^4/\bZ_k$, in turn, can be conveniently computed by taking a
compact Calabi-Yau $X_k=T^8/Z_k$, computing $\chi(X_k)$ and then
dividing the result by the number of fixed points --- taking into
account that the resulting fixed points may be of different
types. This computation was done in \cite{Bergman:2009zh}, with the
result
\begin{equation}
  \chi(\bC^4/\bZ_k) = k - \frac{1}{k}
\end{equation}
or in terms of the M2 charge of the $\text{OM}_{k,0}$ orbifold (the
$\bC^4/\bZ_k$ singularity with no torsion)
\begin{equation}
  Q(\text{OM}_{k,0}) = -\frac{1}{24}\left(k - \frac{1}{k}\right)\, .
\end{equation}
The results are collected in table~\ref{kas}. Adding up the induced
charges of the fixed points on the orbifolded $T^2$ fibration,
according to the fixed point topology described above, we obtain the
D3 charge for the $\OF_{k,0}$ fixed point, as in table~\ref{kas}.

\subsubsection*{Discrete torsion}

In the same way that the O3 plane comes in different flavors depending
on the value of the discrete fluxes, one may expect that the
generalized $\OF$ planes also come in different flavors,
distinguished by discrete flux data.

Consider first the neighborhood of an $\OM_{k,0}$ plane. Since
$H^4(S^7/\bZ_k,\bZ)=\bZ_k$, we can add $p=0,1,\dots,k-1$ units of
discrete torsion to $\OM_{k,0}$ to make an $\OM_{k,p}$. The
contribution to the charge coming from the discrete torsion was
computed in \cite{Aharony:2009fc}, refining the discussion in
\cite{Sethi:1998zk}:\footnote{In the first version of this paper we
  used (with an unfortunate typo) the formulas in \cite{Sethi:1998zk},
  instead of those proposed in \cite{Aharony:2009fc}. Using the
  expression in \cite{Aharony:2009fc} leads to a different
  classification of orientifold types.}
\begin{equation}
  Q(\OM_{k,p}) = Q(\OM_{k,0}) + \frac{p(k-p)}{2k}\, .
\end{equation}

Let us compute the D3 charge of the $\OF_3$ for all choices of
flux. Looking at figure~\ref{fig:tori} we see that the the M-theory
lift is constructed out of three $\OM_3$ points. Each one has a
$\bZ_3$-valued flux associated to it, for a total of 27
possibilities. Out of these some are equivalent: we can choose any of
the three $\OM_3$ planes to lie at the origin of the unit cell, and an
overall reflection of the choice of unit cell does not change the
M-theory geometry. This reduces the number of possibilities to 10. By
enumeration we find that the possible charges in the set are given by
$\{-\frac{1}{3}, 0, \frac{1}{3}, \frac{2}{3}\}$ (with multiplicities,
which we omit).

\begin{table}
  \renewcommand{\arraystretch}{1.25}
  \centering
  \begin{tabular}{c|c}
    Orientifold & Charges \\
    \hline \hline
    $\OF_2$ & $-\frac{1}{4}, 0, \frac{1}{4}, \frac{1}{2}, \frac{3}{4}$ \\
    \hline
    $\OF_3$ & $-\frac{1}{3}, 0, \frac{1}{3}, \frac{2}{3}$ \\
    \hline
    $\OF_4$ & $-\frac{3}{8}, -\frac{1}{8}, 0, \frac{1}{8}, \frac{1}{4}, \frac{3}{8}, \frac{1}{2}, \frac{5}{8}, \frac{3}{4}, \frac{7}{8}$ \\
    \hline
    $\OF_6$ & $-\frac{5}{12}, -\frac{1}{6}, -\frac{1}{12}, 0, \frac{1}{6}, \frac{1}{4}, \frac{1}{3}, \frac{1}{2}, \frac{7}{12}, \frac{2}{3}, \frac{5}{6}, \frac{11}{12}$
  \end{tabular}
  \caption{D3 charges (in F/M-theory conventions, i.e. counting mobile
    branes) for the different $\OF_k$ planes for all possible choices
    of torsion, starting with the configuration with no torsion. We
    have included all possibilities allowed from the M-theory
    perspective, even if some are known or expected to lead to
    unorientifolded IIB backgrounds in the F-theory limit.}
  \label{table:OF3-charges}
\end{table}

The exercise can be repeated for the other $\OF_k$ planes, with the
result shown in table~\ref{table:OF3-charges}. Notice that contrary to
the $\OF_2$ case, we do not necessarily expect that the different
$\OF_k$ planes are distinguished by their D3 charge only (up to
$SL(2,\bZ)$), so there may be more than one inequivalent $\OF_k$ plane
with the same D3 charge. It would be very interesting to clarify this
point, but we keep the notation for convenience in any case.

\subsection{Preserved supercharges}
\label{sec:supercharges}

\begin{table}
\renewcommand{\arraystretch}{1.25}
\begin{center}
\begin{tabular}{|c||c|c|c|c|}
\hline
 & $SU(4)_R$ & $SL(2,\mathbb Z)$ & $SU(2)_L\times SU(2)_R$  &  $\Delta$  \\
\hline\hline
$\phi^I$ & $\mathbf{6}$   & $\mathbf{1}_0$ & $(0,0)$ & $1$ \\
\hline
$\lambda_{\alpha}^a$ & $\mathbf{4}$ & $\mathbf{1}_{\frac{1}{2}}$ & $\left(\frac{1}{2},0\right )$  & $\frac{3}{2}$\\
\hline
$\left (\begin{array}{c} F_{\mu\nu}\\ \star F_{\mu\nu}\end{array}\right )$ & $\mathbf{1}$& $\mathbf{2}_0$ & $\left(1,0\right )\oplus \left(0,1\right )$ & $2$   \\
\hline
$Q_{\alpha a}$ & $\bar{\mathbf{4}}$ & $\mathbf{1}_{\frac{1}{2}}$ & $\left (\frac{1}{2},0 \right )$ & $\frac{1}{2}$ \\
\hline
\end{tabular}
\end{center}
\caption{\small Charges of the different fields on $\mathcal N=4$ SYM in four dimensions.}
\label{table:N=4-reps}
\end{table}

We have made the claim above that the $\OF_k$ planes, for $k=3,4,6$,
preserve twelve supercharges. We now justify this claim from the point
of view of the field theory (we will reproduce the same conclusions
below by analyzing the string construction directly, but it may be
interesting to give a purely field theoretical derivation). We start
with $\cN=4$ $U(N)$ SYM theory. The charges of the fields under the
different symmetries of the theory are shown in
table~\ref{table:N=4-reps}. The subscript in the representation under
$SL(2,\mathbb Z)$ denotes the charge under the $U(1)$ bundle with
transition functions \cite{Kapustin:2006pk,Bianchi:2011qh}
\begin{equation}\label{u1trans}
\gamma=\frac{c\tau +d}{|c\tau +d|}.
\end{equation}

As explained in \S\ref{sec:O3}, an O3 plane is associated with a
quotient by the product of $r$ and $s$. Under $r$, the supercharges
transform as
\begin{equation}
r\colon Q_{\alpha a} \rightarrow -\sqrt{-1} Q_{\alpha a}
\end{equation}
whereas under $s$ they behave as
\begin{equation}
s\colon Q_{\alpha a} \rightarrow \sqrt{-1} Q_{\alpha a}.
\end{equation}
Thus, under the combined action of R-symmetry and $SL(2,\mathbb Z)$
they remain invariant and therefore the quotient preserves sixteen
supercharges, as is familiar.

For higher $k$, we have that the R-symmetry rotation $r_k$ corresponds
to a rotation of $2\pi/k$ along every 2-plane, in the spinorial
representation. Concretely, it is given by (as an element in
$SO(6)_R$)
\begin{equation}
R_k=\left (\begin{array}{ccc}
\hat R_k&0&0\\
0&\hat R_k&0\\
0&0&\hat R^{-1}_k\end{array}\right )
\end{equation}
with $\hat R_k$ a $2\times 2$ matrix corresponding to a $2\pi/k$
rotation. Thus, the action on a spinor of negative helicity (such as
$Q_{\alpha a}$) is
\begin{eqnarray}\nonumber
(-,-,-)&\rightarrow& e^{-i\pi/k}(-,-,-)\\\nonumber
(-,+,+)&\rightarrow& e^{-i\pi/k}(-,+,+)\\\label{negative}
(+,-,+)&\rightarrow& e^{-i\pi/k}(+,-,+)\\\nonumber
(+,+,-)&\rightarrow& e^{3i\pi/k}(+,+,-)
\end{eqnarray}
so its action on the supercharges is
\begin{equation}
r_k: \begin{array}{l}Q_{\alpha A} \rightarrow  e^{-\pi i/k } Q_{\alpha A}\\
Q_{\alpha 4} \rightarrow  e^{3\pi i/k }Q_{\alpha 4}\end{array}
\end{equation}
where the index $A$ runs from 1 to 3. Furthermore, the action of the
S-duality generator $s_k$ is the same on all supercharges
\begin{equation}
s_k: Q_{\alpha a} \rightarrow e^{\pi i/k } Q_{\alpha a}.
\end{equation}
The product $r_k\cdot s_k$ only leaves $Q_{\alpha A}$ invariant, so
the theory preserves twelve supercharges.

\subsection{$\cN=3$ and CPT invariance}

\label{sec:CPT}

The field theories discussed in the previous section have $\cN=3$
supersymmetry in four dimensions. This is surprising since there is a
well known argument \cite{Weinberg:2000cr} which states that, in the
absence of gravity, perturbative four dimensional theories with twelve
supercharges are actually $\cN=4$. We review such argument in the
following stressing the fact that it only applies to theories with a
perturbative limit. Since the cases we are studying are intrinsically
non-perturbative, the argument does not apply. Furthermore, we will
explicitly show that the operator spectrum of these theories does not
fall into representations of $SU(4)_R$, so there is no possibility for
the supersymmetry to enhance to $\cN=4$ (which could have happened in
principle, even if the standard argument for enhancement does not
apply).

Consider an $\cN=3$ supersymmetric QFT in 4d Minkowski space. This
means that the $\cN=3$ supersymmetry algebra has a well-defined action
on the space of operators that define the theory, so these can be
arranged in irreducible representations of the supersymmetry
algebra. As usual, we want CPT to be a symmetry of the theory so it
should also act on the space of operators. Then, under CPT an
irreducible representation of supersymmetry may be mapped to itself or
it can happen that two of them are mapped to each other.

Now let us assume that the theory has a Lagrangian description. This
means that there are certain preferred operators (elementary fields)
which are the ones that appear in the Lagrangian and in the measure of
the path integral. The Lagrangian may depend on some parameters and we
assume there is a point in such parameter space where the theory
becomes free (the Lagrangian is quadratic in the elementary
fields). At that point, the operator algebra can be recovered from the
elementary fields and the Hilbert space obtained from canonical
quantization. There are various free theories that one can consider
at this point:

\begin{itemize}

\item The elementary fields include only massless vector
  multiplets. As it happens, there are two kinds of $\cN=3$ vector
  multiplets that get mapped into each other under CPT and the two of
  them together make a vector multiplet of $\cN=4$. All the possible
  actions involving these fields turn out to be invariant under
  $\cN=4$ \cite{Weinberg:2000cr}. Notice that since the whole space of
  operators is built from the elementary fields (which are in an
  $\cN=4$ representation) we have that the action of the $\cN=4$
  algebra extends to all of them. Now one can argue that since the
  number of supersymmetries is discrete, it cannot change as we move
  continuously towards the strong coupling regime. This is the
  argument that in gauge theories, $\cN=3$ supersymmetry is actually
  $\cN=4$.

\item The next option is to include a gravity multiplet (and vector multiplets possibly). Again, there are two multiplets which get mapped into each other so we need to include them both at the same time. However, in this case the direct sum of the two does not admit the action of the full $\cN=4$ algebra. Therefore, we find genuine $\cN=3$ (supergravity) theories.

\end{itemize}

In a non-perturbative theory, there is no argument that suggests that
there is an enhancement from $\cN=3$ to $\cN=4$ for non-gravitational
theories which preserve CPT. Then, if there is an operator in a
representation of $\cN=3$ which cannot combine with others to form a
representation of $\cN=4$, the theory only has $\cN=3$. As we
argue in the following, this can be explicitly checked for the
theories we are considering and there are indeed operators which only
admit the action of $\cN=3$.

For simplicity, we start with the free theories obtained as a $\bZ_k$
quotient of $U(1)$ $\cN=4$ SYM. The chiral primaries in the parent
theory correspond to symmetric traceless products of $n$ scalars
$\phi^I$, which has conformal dimension 1 (see for instance
\cite{Aharony:1999ti}). Let us look at the case $n=2$ and check which
operators survive the quotient. We have that
$\mathbf{6}\times\mathbf{6}=\mathbf{1}\oplus\mathbf{15}\oplus\mathbf{20'}$
where the symmetric traceless piece corresponds to the $\mathbf{20'}$.
This is the only chiral primary of conformal dimension $\Delta=2$.

Since $\bZ_k$ (for $k>2$) breaks the R-symmetry $SU(4)_R\rightarrow S(U(3)\times U(1)_P)$, we split the scalar fields as follows,
\begin{equation}\label{asd}
\mathbf{6}\rightarrow \mathbf{3}_{1,1}\oplus\mathbf{\bar 3}_{-1,-1}.
\end{equation}
The first subindex denotes the charge under $U(1)_P$ (in some normalization) and the second is the charge under $\bZ_k$ (which is an integer mod $k$). All we have to do now is take the $\mathbf{20'}$ we had in the original theory, decompose it as in \eqref{asd} and keep only the states invariant under $\bZ_k$. This gives
\begin{equation}
\mathbf{20'}\rightarrow \mathbf{8}_{0,0}\oplus\mathbf{6}_{2,2}\oplus\mathbf{\bar 6}_{-2,-2}\,,
\end{equation}
so the last two factors are not invariant under $\bZ_k$ (for $k>2$)
and are projected out. The remaining representation $\mathbf{8}_{0,0}$
clearly does not admit the action of $SU(4)_R$ so we cannot have an
enhancement to $\cN=4$.

This argument is in fact also valid for interacting theories since the
conformal dimension is protected for chiral primaries, so we conclude
that there is no enhancement to $\cN=4$ for any of the theories with
$k>2$.

As a final comment, notice that the fact that the marginal operator
associated with changes in the Yang-Mills coupling is projected out
can also be understood from this viewpoint. The relevant marginal
operator is of the form $F\wedge \star F + i F\wedge F+\ldots$, given
by a component with conformal dimension 4 in the $n=2$ multiplet. It
is a singlet of $SU(4)_R$, but transforms nontrivially under
$SL(2,\bZ)$ for $k>2$ (for instance, for $k=4$ we have
$(F,\star F)\to (\star F, -F)$, so the multiplet gets an overall minus
sign), so it gets projected out in the quotient $\cN=3$ theory.

\subsection{Large $N$ limit}

\label{sec:large-N}

One can place an arbitrary number of D3 branes on top of the $\OF$
planes, so a natural question is whether the system admits a dual
holographic description at large $N$. The orbifold description of the
theory suggests that the answer is positive: start with the
holographic dual of $U(N)$ $\cN=4$ SYM, and quotient by
$r_k\cdot s_k$. The $r_k$ generator becomes a freely acting $\bZ_k$
action on the $S^5$, while $s_k$ maps to the $SL(2,\bZ)$ duality of
IIB string theory, as usual.

As in previous cases, the $k=2$ case \cite{Witten:1998xy} may be
illuminating. We have that the orientifold projection maps to the
$\bZ_2$ involution $\sigma\colon S^5\to \bR\bP^5$. In addition to the
geometric action, it acts with $(-1)^{F_L}\Omega$ on the theory, which
is encoded in a reflection of the torus fiber, if we represent the
background in F-theory.

An equivalent but perhaps clearer description is to start with the
F-theory description of the system, as $AdS_5\times S^5 \times T^2$,
and taking a freely acting quotient by a $\bZ_k$ symmetry acting
simultaneously on the $S^5$ and $T^2$ factors.

Either way, it is clear that the axio-dilaton of the the holographic
system will be projected out, in accordance with the fact that the
dual $\cN=3$ theory on the boundary has no marginal deformation
associated with the complexified gauge coupling. This fact makes the
dual gravity description somewhat subtle, but it may still be
approachable in the M-theory picture.

\medskip

It is also interesting to compute the amount of supersymmetry
preserved by the holographic dual. We do so by first looking at $N$ D3
branes probing an $\OF_k$ and then taking the near horizon limit. The
computation runs parallel to that in \S\ref{sec:supercharges}, and it
is in fact more standard (see for example
\cite{Font:2004et,Aharony:2008ug}), so we will be brief.

Type IIB has two supercharges $Q_{10}^+$ (Weyl spinors of positive helicity) in
ten dimensions, which are decomposed under $SO(1,9)\rightarrow SO(1,3)\times SO(6)$
as
\begin{equation}
Q_{10}^+=(Q_4^+\otimes Q_6^+)\oplus(Q_4^-\otimes Q_6^-),
\end{equation}
where the first (second) term is a positive (negative) helicity Weyl
spinor in both four and six dimensions. As explained earlier, $r_k$
corresponds to an R-symmetry rotation, which is realized in the ten
dimensional perspective as a rotation in the six dimensions transverse
to the $\OF_k$.  Thus, we have that $Q_6^-$ transforms as
\eqref{negative} and analogously for $Q_6^+$.  Furthermore we have
that $Q_6^\pm$ has charge $\pm \frac{1}{2}$ under the S-duality $U(1)$
bundle \eqref{u1trans} so we find that $s_k$ is given by
\begin{equation}
s_k: Q_6^\pm \rightarrow e^{\pm\pi i/k } Q_6^\pm\,,
\end{equation}
and under the combined action $r_k\cdot s_k$ only twelve supercharges survive for $k>2$.
Here we see again that it is crucial to include a non-trivial action under $SL(2,\mathbb Z)$ to
preserve any supersymmetry.

Finally, the presence of $N$ parallel D3 branes does not break
supersymmetry further so the full system is indeed $\mathcal N=3$ from
a four dimensional viewpoint. This is still true once we take the near
horizon limit which gives Type IIB on $AdS_5\times (S^5/\mathbb Z_k)$
with a non-trivial $SL(2,\mathbb Z)$ bundle over
$S^5/\mathbb Z_k$.\footnote{In \cite{Ferrara:1998zt} a
  construction of $\cN=6$ supergravity in $AdS_5$ similar to the one
  proposed here was described.}

\section{Conclusions}

\label{sec:conclusions}

In this paper we have argued for the existence of interesting
nontrivial theories arising from D3 branes probing what are
essentially non-perturbative F-theory generalizations of the O3
plane. For the cases that we have studied, the existence of these
generalized orientifold planes projects out the axio-dilaton, so the
resulting theories have no marginal deformations associated to the
coupling.

The simplest case to study, beyond the well understood O3 plane, are
the $\OF_k$ planes with $k\in\{3,4,6\}$. We have seen that they
preserve twelve supercharges, providing the first examples (beyond
supergravity) of $\cN=3$ theories in four dimensions. These theories
have to be necessarily somewhat exotic in order to avoid well-known
theorems about enhancement to $\cN=4$, and we have shown how indeed it
seems to be the case that our theories evade the assumptions of the
no-go results.

The F-theory construction provides an intuitive way of understanding
these field theories as exotic ``S-duality orbifolds'' of ordinary
$\cN=4$ theories, which explain readily some of their most puzzling
properties. When it comes to further developments, it may be easier to
study the string theory realization of the theories instead, and in
this paper we have given some concrete steps in this direction,
providing an explicit M-theory realization.

\medskip

\subsection*{Further directions}

Our analysis has been focused on arguing for the existence of these
theories, and finding some of their most elementary properties. There
are clearly a large number of further avenues of study, we will
highlight here a few.

\medskip

A first observation is that if we compactify the system on a circle,
it admits a dual description as a stack of M2 branes moving on a
background with singularities of the type $\bC^4/\bZ_k$. So the
compactified theories flow at low energies to ABJM
\cite{Aharony:2008ug,Aharony:2008gk} at certain loci of their moduli
spaces. We should thus be able to gain quite a bit of insight into
these $\cN=3$ theories by studying their flow to the well understood
(by comparison) ABJM theories. As an example, one can hope to gain
information about the four dimensional theories by studying the
superconformal index of appropriate ABJM theories. The $k=1,2$ cases
have in fact been approached in a related way \cite{Gang:2011xp}, and
it would be rather interesting to generalize this analysis in order to
learn more about the class of theories introduced in this note.

\medskip

Also, we have been strongly guided by the IIB string construction, but
this comes at the risk of missing possible consistent theories. It
would be desirable to sharpen the purely field theoretical description
of our construction, in order to have a purely field theoretical
understanding of which orbifolds are allowed and which extra massless
particles one must include for each choice of field theory
orbifold. For instance: we started with $U(N)$ theories, coming from
the D3 branes, but this is certainly not the only known example of
$\cN=4$ theories. Perhaps other $\cN=3$ theories, beyond the ones
discussed here, can be constructed starting from $\cN=4$ theories with
other gauge groups. The harmonic superspace formulation of $\cN=3$ SYM may be
helpful in this regard \cite{Galperin:1984bu,Galperin:1985uw,Delduc:1988cp,Ivanov:2001ec,Buchbinder:2004rj,Galperin:2007wpa,Buchbinder:2011zu}.

\medskip

Conversely, the string picture clearly shows that there are other,
less supersymmetric, generalized orientifold planes that one can
construct in F-theory. There are certainly a number of less
supersymmetric M-theory orbifolds we could have taken, and we could
also try to study other non-orbifold elliptic fibrations with complex
codimension four singularities. Often it will not be possible to
deform or resolve the resulting geometries into a neighboring smooth
Calabi-Yau, so they are out of reach of conventional F-theory
techniques (with possibly the remarkable exception of
\cite{Collinucci:2014taa}). Nonetheless the close connection
between these generalized orientifolds and ABJM-like theories --- a
field in which much progress has been achieved in the last years ---
gives a promising window into this interesting class of constructions.

\acknowledgments

We thank Anamaría Font, Tomás Gómez, Jan Keitel, Marina Logares and
Ángel Uranga for enlightening discussions, and especially Ben
Heidenreich for helpful comments on the draft. D.R is supported by a
grant from the Max Planck Society.

\appendix

\section{Self-duality of O3$^-$}

\label{ap:self}

On the one hand, we claim that the theory associated to $N$ physical
D3-branes probing an O3$^-$ has gauge group $O(2N)$ rather than
$SO(2N)$. On the other, there are arguments suggesting that such a
theory should map to itself under S-duality (but with different
coupling generically). It is well known that the theory with gauge
group $SO(2N)$ is indeed self-dual and we would like to show in the
following that the same is true for $O(2N)$. We do so for the case
$N=1$, since it corresponds to a free theory and the duality can be
performed explicitly at the level of the path integral (see for
instance \cite{KapustinLectures}). However, we expect the same to be
true also for arbitrary $N$, we give a heuristic argument for this
below.

The partition function of the theory (in Euclidean signature) is given
by
\begin{equation}
Z=\sum_E\int \mathcal D \phi\,\mathcal D \lambda\,\mathcal D A\, e^{-S_E}
\end{equation}
where $S_E$ is the classical action for $\mathcal N=4$ SYM with gauge
algebra $\mathfrak{so}(2)$. The global structure of the gauge group
enters in two different places. First, in the sum over topologically
distinct gauge bundles $E$. For gauge group $O(2)$, one must sum over
all bundles with transition functions valued in $O(2)$ which are, in
general, not in $SO(2)$.\footnote{When the spacetime is
  simply-connected we actually have that every $O(2)$ bundle is
  equivalent to an $SO(2)$ bundle \cite{PhysRevD.17.3196}. However, we will also be interested
  in the theory on $\mathbb R^3\times S^1$ where this becomes
  important. See for instance the comment in footnote~\ref{fn:O(2)}.}
Second, as explained in the main text, the $\mathbb Z_2$ in $O(2)$
which is not in $SO(2)$ acts on the different fields with a minus
sign. Thus, two field configurations related by such $\mathbb Z_2$
should be considered equivalent and only included once in the path
integral.

Since the theory is free, the dynamics of the scalars, fermions and vectors decouples and we can focus just on the duality for the vectors, which already shows the relevant point. Thus, consider the path integral 
\begin{equation}\label{partition}
\hat Z=\sum_E\int \mathcal D A\, e^{-\hat S_E}
\end{equation}
with
\begin{equation}
\hat S_E = \int_{\mathbb R^4} \frac{1}{2e^2}F\wedge \star F-\frac{i\theta }{8\pi^2}F\wedge F
\end{equation}
where we took spacetime to be $\mathbb R^4$ for simplicity. Notice that $\hat S_E$ is invariant under the action of $\mathbb Z_2$ which flips the sign of the field strength $F$. Following \cite{KapustinLectures} we can rewrite this integral as
\begin{equation}
\hat Z=\int \mathcal D F\,\mathcal DB\, e^{-\hat S_E+i\int B\wedge dF}
\end{equation}
where the integration over the one-form $B$ is included so that once we integrate over it we effectively restrict ourselves to closed field strengths. For the case in which spacetime in $\mathbb R^4$, this is indeed equivalent to \eqref{partition}. If, on the other hand, we integrate over $F$, we end up with the dual description of the theory in which $B$ is the gauge potential and the gauge coupling is $-1/\tau$. The crucial point is that in order to make the extra term $B\wedge dF$ invariant under $\mathbb Z_2$, we must declare that $B$ is odd under it. Thus, we see that that the $\mathbb Z_2$ that acts on the electric variables must act \emph{at the same time} on the magnetic ones. In other words, the dual theory is again given by an $O(2)$ theory rather than $SO(2)$. This can also be seen from the realization of the $\mathbb Z_2$ as the product of R-symmetry and $SL(2,\mathbb Z)$, as explained in section \ref{sec:O3}.

There is a slightly different way to see this which applies to $O(2N)$ for any $N$. One can regard the $O(2N)$ theory as arising from a $\bZ_2$ quotient of $SO(2N)$, where the $\bZ_2$ acts on both the electric and magnetic descriptions in the same way. Thus, since the original $SO(2N)$ theory is self-dual, we expect the same is true for $O(2N)$.

\bibliographystyle{JHEP}
\bibliography{refs}

\end{document}